\documentclass[final,1p,times,authoryear]{elsarticle}
\usepackage{graphicx}%
\usepackage{multirow}%
\usepackage{float}
\usepackage{amsmath,amssymb,amsfonts}%
\usepackage{amsthm}%
\usepackage{mathrsfs}%
\usepackage[title]{appendix}%
\usepackage{xcolor}%
\usepackage{textcomp}%
\usepackage{manyfoot}%
\usepackage{booktabs}%
\usepackage{algorithm}%
\usepackage{algorithmicx}%
\usepackage{algpseudocode}%
\usepackage{listings}%
\usepackage{hyperref}
\usepackage{comment}
\hypersetup{
	colorlinks=true,
	linkcolor=cyan,
	filecolor=blue,      
	urlcolor=red,
	citecolor=green,
}

\usepackage{amssymb}
\usepackage{lineno}

\journal{Measurement}

\begin{document}

\begin{frontmatter}


\title {Depth from Defocus Technique for High Number Densities and Non-spherical Particles }



\author[label1,label2]{Rixin Xu}
\author[label1,label2]{Zuojie Huang}
\author[label1,label2]{Wenchao Gong}
\author[label1,label2]{Wu Zhou\corref{cor1}}
\ead{zhouwu@usst.edu.cn}
\cortext[cor1]{Corresponding author at: School of Energy and Power Engineering, University of Shanghai for Science and Technology, Shanghai, 200093, China}
\author[label3]{Cameron Tropea}
\affiliation[label1]{organization={School of Energy and Power Engineering, University of Shanghai for Science and Technology},
            city={Shanghai},
            country={China}}
\affiliation[label2]{organization={Shanghai Key Laboratory of Multiphase Flow and Heat Transfer for Power Engineering, University of Shanghai for Science and Technology},Department and Organization 
             city={Shanghai},
            country={China}}
\affiliation[label3]{organization={Institute for Fluid Mechanics and Aerodynamics, Technical University of Darmstadt}, city = {Darmstadt}, country={Germany}}

\begin{abstract}
The Depth from Defocus (DFD) imaging technique for measuring the size and number concentration of particles in a dispersed two-phase flow has up to now been restricted to relatively sparse particle densities and to identifying only spherical particles. The present study examines two advancements to the technique, widening its range of application significantly. The first advancement introduces an image processing procedure which can identify and size particle images which are overlapping. This increases the tolerable number concentration of particles which can be identified and processed within the measurement volume. The second advancement explores the possibility of determining the size and position of non-spherical particles within an observation volume. Both advancements  build on recent theoretical work,  utilizing not only the gray level of the blurred, out-of-focus images, but also the gradient of the gray level normal to the nominal particle  or particle ensemble contour.  This gray-level gradient is used to  estimate the width of the blur kernel, which is assumed to remain Gaussian.

These enhancements  are experimentally validated using a dedicated apparatus in which particles of known size, shape and degree of overlapping images can be systematically varied. This experimental setup provides benchmark data to quantify the accuracy and limitations of the processing algorithms.
\end{abstract}


\end{frontmatter}

\section{Introduction}
The technique of obtaining depth and size information from defocused blurry images of particles is known as Depth from Defocus (DFD). This concept was initially proposed in \cite{pentland1987new}  and has since become one of the popular depth estimation techniques, parallel to stereo vision and holographic imaging. This method can be used to determine the size and location of spherical particles in a fluid, even when their position is beyond the depth of field limit; thus, out of focus. Moreover, this  allows all particles within a well-defined volume to be counted and sized,  allowing accurate measurement of number and volume concentration. 

The DFD principle can be implemented through various optical means to obtain defocused images, including changing aperture between two exposures, thus  inducing a controlled defocus in the formed images. Alternatively, the technique has been realized using either  one or two cameras. In both cases the goal is to identify and size particles over a three-dimensional volume, while also determining their coordinates. With appropriately fast cameras, it is  possible to resolve the trajectory of the particles in time using particle tracking velocimetry \citep{willert1992three,murata1999particle,bao2011defocus}, although this aspect will not be elaborated further in the present study. 

First defocus systems for particle measurement used two images, varying the defocus blur by adjusting imaging system parameters between images, such as aperture, focal length, and the distance from the lens to the camera imaging plane \citep{subbarao1988depth,surya1993depth}. In the study by \cite{lebrun1994simultaneous}, a beam splitter was employed to evenly divide the imaging rays into two cameras. By adjusting the spacers connecting the dual camera setup, it became possible to simultaneously capture two images of particles with different degrees of blur at a specific moment, with one camera. The disparity in blur between these two images was used to assess defocus ambiguity. In terms of image processing, various methods were employed in the early stages, including spatial domain deconvolution restoration \citep{ens1991matrix} and frequency domain processing, using a Fourier transformation \citep{zhou2020estimation}, to identify the degree of blur in particle images, corresponding to their depth position. More recent work involving two cameras  has the advantage that it can distinguish whether a particle is in front of or behind the object plane of the imaging system \citep{zhou2020spray}. The sensitivity of this technique to a large number of optical and system realization parameters has been investigated in \cite{zhou2021sensitivity} and has been used for shock-drop interaction studies in \cite{sharma2023depth}.

More attractive in terms of equipment complexity is the one camera realization, realized in a number of configurations in the past \citep{subbarao1988parallel,cierpka2010simple}. One particular configuration uses a cylindrical lens in the imaging optics, such that the out-of-focus condition results in an astigmatism to determine the position of the particle along the optical axis \citep{barnkob2015general,barnkob2020general}. Recently, neural networks have been adopted to deal with these out-of-focus images \citep{barnkob2021defocus, zhang20233d}. In addition, there is a method for particle depth position measurement by employing specially shaped aperture configurations, primarily including three-hole apertures and annular apertures \citep{willert1992three,pereira2002defocusing,levin2007image}. The arrangements of these apertures allow for determining whether particle coordinates are in front of or behind the object plane. On the other hand, the emphasis of these studies has not been placed on also determining the size of the particles.

The following study concerns the one camera approach to realize the DFD technique and relies on extracting the size and position coordinates of the particle from the blurred image, without any astigmatism introduced into the system. This means that the blurring is close to identical, whether the particle is in front of or behind the object plane. Thus, the present system cannot differentiate between these cases. This is not necessarily a handicap, since the object plane can be  positioned such that the entire volume of interest can either be in front of or behind the plane.

The technique to be used in the present study  for analysing images of both   particles with overlapping images and non-spherical particles builds on recent work introduced in \cite{jatin2024depth}. In this approach, not only the blurred gray levels are used, but also their local gradient in a direction normal to the image contour. From this gray level gradient, an estimate of the blur kernel width is made. This eliminates the need for calibration of the technique when determining size, although a calibration is still required for determining the particle coordinate along the axis of the imaging system ($z$ direction).


First, in section~\ref{sec:gradient} the procedure for sizing and locating spherical particles using the gray level gradient is briefly summarized, as this builds the basis for this particular  one-camera realisation of the technique. A situation is then examined, arising when neighbouring particles result in overlapping projected images, for instance when measuring  high number density dispersions or if agglomerates exist in the dispersion (subsection~\ref{subsec:overlap}). An attempt to determine number concentration limits is made, in which the degree of overlap is systematically varied, while monitoring the accuracy of the size and location estimates (section~\ref{subsec:validation multiple}).

 The next step is to introduce a procedure for analysing images of non-spherical particles (subsection~\ref{subsec:Non_Spherical Particles}). For this, the contour of the particle image at a prescribed gray level is first determined and then the local gray level gradient along the normal to this contour is found. The contour is used to compute the diameter of an area equivalent circle and this is used in the conventional manner to determine the size and location of a corresponding spherical particle. The scaling ratio between the image size and estimated true particle size is then applied to re-scale the non-spherical contour, yielding an estimate of the contour of an in focus, non-spherical particle. This algorithmic approach is then applied to well controlled  laboratory images to quantify its accuracy and limitations (subsection~\ref{subsec:validation non-spherical}).


Having successfully extended the DFD technique to the measurement of dispersions at high concentrations and non-spherical particles, the study concludes with a discussion of new realms of application and the challenges that remain to be met in implementation (Section~\ref{sec:Conclusions}).

\section{Analytic Relations for a Single Camera DFD}
\label{sec:gradient}
\begin{figure*}[ht]
\centering
\includegraphics[width=1\linewidth]{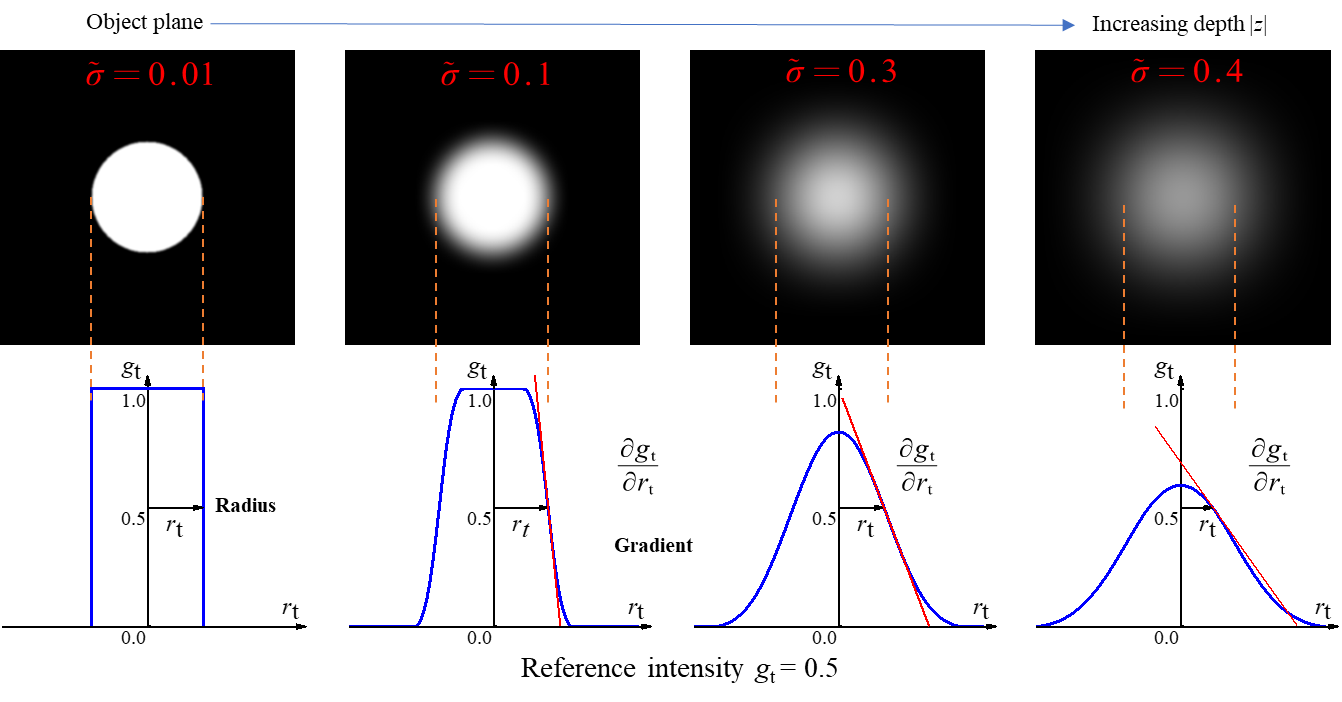}
\caption{Illustration of particle image and resulting intensity distribution for varying degrees of out-of-focus. Two quantities are extracted from the images – radius ($r_\mathrm{t}$) and intensity gradient ($\partial g_\mathrm{t}/\partial r_\mathrm{t}$) at a reference intensity value ($g_\mathrm{t}$=0.5); both of which decrease with increasing depth from the object plane. $\Tilde{\sigma}$ is a dimensionless $\sigma$ defined in Eq.~(\ref{eq:basic-nondim}). }
\label{fig:blurring}
\end{figure*}

This description will intentionally be kept brief, because it simply summarizes the work in \cite{jatin2024depth}.  Typical images from in-focus and out-of-focus particles are illustrated in  Fig.~\ref{fig:blurring}.   From this figure, it is evident that the degree of image blur depends on the out-of-focus distance $z$ from the object plane. However, also the gradient of the gray level changes with $z$. These blurred images can be described by  a convolution of the focused image $f(x,y)$ of a particle  with a blur kernel $h(x,y)$ \citep{blaisot_droplet_2005, zhou2020spray}. The intensity $g_\mathrm{t}$ at any location on the sensor plane $(x,y)$ is then evaluated as:
\begin{equation} \label{eq:convolution}
g_\mathrm{t}(x,y) = f(x,y) \circledast h (x,y)   
\end{equation}
where $f(x,y)$ is a normalized intensity  of the in-focus particle  on the image plane  
\begin{equation} \label{eq:f_r}
\begin{split}
f(x,y) & = 0,    \; \textrm{if}    \;   \mathrm{outside \;  particle\; contour} \\
& = 1,         \;    \textrm{if}    \; \mathrm{inside\; particle\; contour} 
\end{split}
\end{equation}
The particle dimensions on the image plane are related to the actual size  through the magnification factor of the optical system, $M$.  
The blur kernel $h(x,y)$ can be approximated by a Gaussian function with $\sigma$ as the standard deviation \citep{junjie2023image}:
\begin{equation} \label{eq:h_r}
h(x,y) = \frac{1}{2 \pi \sigma^2} e^{-{\frac{x^2+y^2}{2\sigma^2}}} 
\end{equation}
This assumes that the point spread function has a central peak much narrower in width than $\sigma$, which is fulfilled in typical DFD optical systems \citep{jatin2024depth}. The standard deviation $\sigma$ represents the degree of blur or size of the blur kernel and is proportional to the displacement of the particle away from the object plane ($\Delta z$), 
\begin{equation} \label{eq:sig-delz}
\sigma= \beta\lvert z \rvert
\end{equation}
where $\beta$ describes the proportionality and is constant for a given optical system. Furthermore, we assume   that telecentric lenses are used on both the illumination and receiving sides of the optical system; hence, the magnification does not change with particle $z$ position. 

Given this description of the blurred images, there are several avenues to follow with the aim of determining the original image from the defocussed blurred image $g_\mathrm{t}(x,y)$. Conventional approaches use a non-blind deconvolution. This is non-blind since the blur kernel is known, arising from Gaussian defocus blurring and the point spread function (which in practical systems is negligibly small). The most common non-blind deblurring methods employ the Wiener filter (\cite{wiener1949extrapolation}) or the Richardson-Lucy algorithm (\cite{richardson1972bayesian}). More recently, the performance of these conventional approaches has been significantly enhanced by  integrating them with learned deep features (\cite{dong2021dwdn}). Although these approaches are capable of reconstructing the high resolution, deblurred images, the location of the particle on the $z$ axis is not available. For this reason, a different approach is used in the present study in that the convolution integral is explicitly solved for spherical particles. Having this, and a quantitative calibration of the  parameter $\beta$, the particle $z$ position can be found.

For spherical particles the convolution integral can be solved and using the following dimensionless variables
\begin{equation} \label{eq:basic-nondim}
\widetilde{\rho}=\frac{r}{d_o},\ \ \widetilde{\rho_\mathrm{t}}\ =\frac{r_\mathrm{t}}{d_o},\ \ \widetilde{\sigma}=\frac{\sigma}{d_o}
\end{equation}
the integral can be expressed in dimensionless form as 
\begin{equation} \label{eq:nondim-g_rt}
g_\mathrm{t}\left(\widetilde{\rho_\mathrm{t}}\right)=\frac{1}{{\widetilde{\sigma}}^2}\int_{0}^{1/2}{e^{-\frac{{\widetilde{\rho}}^2+{{\widetilde{\rho}}_\mathrm{t}}^2}{2{\widetilde{\sigma}}^2}}I_o\left(\frac{\widetilde{\rho}\widetilde{\rho_\mathrm{t}}}{{\widetilde{\sigma}}^2}\right)\widetilde{\rho}d\widetilde{\rho}}
\end{equation}
\\
where $I_o$ is the zeroth order modified Bessel function of the first kind and $d_0$ is the true particle diameter.

From the particle image, two quantities are extracted – the radius $\left(r_\mathrm{t}\right)$ and intensity gradient $\left(\partial g_\mathrm{t}/\partial r_\mathrm{t}\right)$ at a reference intensity value, e.g., $g_\mathrm{t}=0.5$. Both of these quantities decrease as the particle is further displaced  from the object plane ($z=0$) (Fig.~\ref{fig:blurring}). This suggests that the gray level gradient can be used to estimate the parameter $\sigma$ in the blur kernel. In \cite{jatin2024depth} a  novel measurable dimensionless radius is proposed \citep{Jatin2024patent}: 
\begin{equation} \label{eq:nondim-rad}
{\widetilde{R}}_\mathrm{t}=\left(\frac{{\widetilde{\rho}}_\mathrm{t}}{\left({\widetilde{\rho}}_\mathrm{t}\right)_{g_\mathrm{t}=0.5}\ }\right)_{\widetilde{\sigma}}=\left(\frac{r_\mathrm{t}}{\left(r_\mathrm{t}\right)_{g_\mathrm{t}=0.5\ }}\right)_{\widetilde{\sigma}}
\end{equation}
where $\left(r_\mathrm{t}\right)_{g_\mathrm{t}=0.5}$ is the radius at the reference intensity. The proposed functional form of the analytic function  is
\begin{equation} \label{eq:calfun2}
\widetilde{G}\ =\left|\frac{\partial g_\mathrm{t}}{\partial{\widetilde{R}}_\mathrm{t}}\right|_{g_\mathrm{t}=0.5}=\left|r_\mathrm{t}\frac{\partial g_\mathrm{\mathrm{t}}}{\partial r_\mathrm{t}}\right|_{g_\mathrm{t}=0.5}=f_2\left(\widetilde{\sigma}\right)
\end{equation}
From this  dimensionless version of intensity gradient $|\partial g_\mathrm{t}/\partial {\widetilde{R}}_\mathrm{t}| = |r_\mathrm{t} \partial g_\mathrm{t}/\partial r_\mathrm{t}|$ at the reference intensity (subscript $g_\mathrm{t}=0.5$ is omitted  from now on), we can estimate the dimensionless depth, expressed as a dimensionless standard deviation of the blurring  $\widetilde{\sigma}$. Note that in \cite{jatin2024depth}, a limiting value of $\widetilde{\sigma}=0.35$ was recommended, below which the size measurement would be  reliable.

Another   function is  necessary  to estimate ${\widetilde{\rho}}_\mathrm{t}$ from $\widetilde{\sigma}$ at the reference intensity, represented in the functional form as  
\begin{equation} \label{eq:calfun1}
{\widetilde{\rho}}_\mathrm{t}=f_1(\widetilde{\sigma})
\end{equation}
These functions can be further combined in the form ${\widetilde{\rho}}_\mathrm{t}=f_1(f_2^{-1}(\widetilde{G}))$. So the particle diameter $d_o$ can be measured after obtaining $\widetilde{G}$ and $r_\mathrm{t}$ from particle image.

The calibration is necessary not for the size determination, but for  the position, i.e., $\beta$ in Eq.~(\ref{eq:sig-delz}). This calibration consists of moving a reticle target along the $z$ axis while registering the standard deviation $\sigma$, using the function ${\widetilde{\sigma}}=f_2^{-1}(\widetilde{G})$ and the true particle diameter $d_o$. 


\section{Image Processing Algorithms}
\label{sec:algorithms}

 \subsection{Procedure for  Spherical Particles with Overlapping Images}
\label{subsec:overlap}
We begin by reviewing the  procedure   for circular images arising from non-overlapping spherical particles. 
For a given normalized gray level ($g_\mathrm{t}$) a Wiener filter is firstly applied to reduce the white noise component. Then the contour of the image is established. This is performed using a bilinear interpolation, achieving a subpixel resolution. The image is then binarized with threshold 0.5, making all pixels within the contour 1 and all outside 0. Working from this modified image, the diameter of the circle is computed. 

Along the threshold contour the direction of the normal vector pointing outward is found and the gray-level gradient along this vector at the boundary is interpolated and  averaged over the entire contour circumference. However, according to Eq.~(\ref{eq:f_r}), the sought gradient assumes a background level of zero (i.e., the background level is completely white), whereas in practice a non-zero background level exists. Thus, the computed gradient must be scaled with a factor related to the background gray level in the vicinity of the particle image. This adjusted gradient is then used with function $f_2$ to determine the blur kernel parameter $\Tilde{\sigma}$ and further determine the size and position of a sphere in the object plane leading to this  diameter on the image plane. 

If now several spherical particles are in close proximity to one another, i.e., high number/volume concentration, then they will generate blurred images which are overlapping on the image plane. To analyse these images, the above algorithmic procedure is extended. First the contour of the overlapping particle images is determined at a selected $g_t$ level, using a bilinear spline interpolation to achieve sub-pixel resolution. An ellipse is fitted to the $g_t$ contour, determining a major and minor axis. Two criteria are then tested before proceeding.  If the ratio of major to minor axis is less than 1.1, then the image is considered to be generated by a single particle and the standard procedure for analysing a single particle is invoked. 

\begin{figure}
    \centering
    \includegraphics[width=1\linewidth]{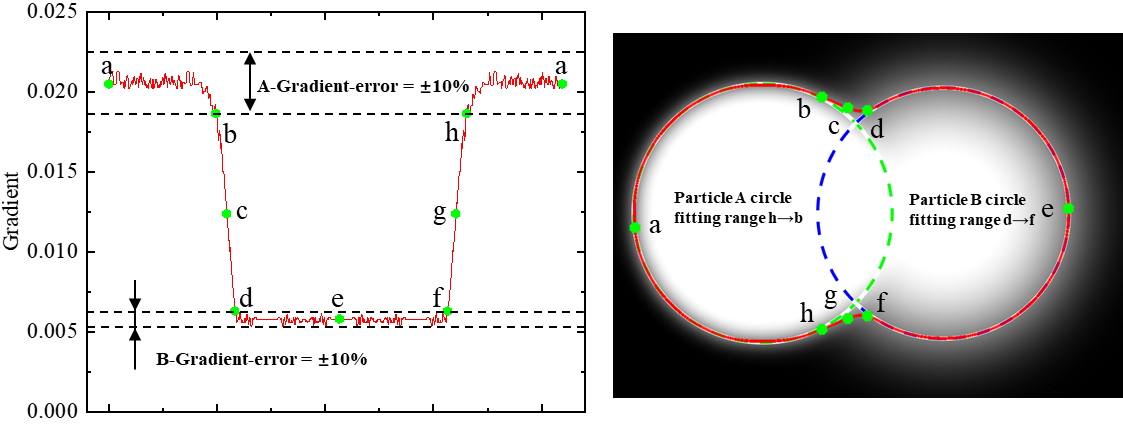}
    \caption{Normalized  gray-level gradient computed from an image arising from two neighbouring particles with overlapping, blurred images. The two example particles have different $z$ positions. }
    \label{fig:gradient overlapping particles}
\end{figure}

The second criterion is illustrated with Fig.~\ref{fig:gradient overlapping particles}, showing  exemplary overlapping images of  two neighbouring particles. The normalized  gray-level gradient around the circumference of the images is also shown. The two end points of the fitted ellipse are marked as points \textbf{a} and \textbf{e} in the  image. Some smoothing of the gradient curve is performed before proceeding. First a representative value of the gray scale gradient at each   end point is computed by averaging the gradient over the neighbouring 10 pixels. Then the  gray-level gradient is averaged over a portion of the $g_\mathrm{t}$-contour over which the values lie within $\pm 10\%$ of the representative value at the end points \textbf{a} and \textbf{e}. These portions of the contour used for averaging are marked on the graph of Fig.~\ref{fig:gradient overlapping particles}. 

Once the two  gray-level gradient values and the associated blur kernel parameter $\tilde \sigma$ for each are determined, the diameter (or radius $r_\mathrm{t}$) of each particle image at $g_\mathrm{t}=0.5$ is computed, assuming circularity of the image and fitting (least squares) a circle to the portion of the $g_\mathrm{t}$-contour used for the above  gray-level gradient averaging procedure. These two values of each particle, $r_\mathrm{t}$ and $\Tilde{G}$, are then used to estimate the true sizes of the two particles, and using the calibration constant $\beta$, the $z$ positions of the particles are found. Note that the calibration parameter $\beta$ is the same for each particle, only the blur kernel parameter $\tilde \sigma$  may differ.

\begin{figure}[h]
    \centering
    \includegraphics[width=0.2\linewidth]{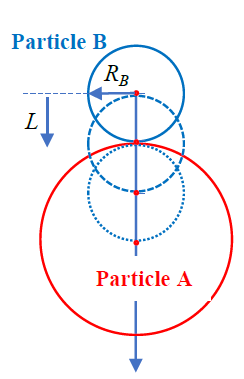}
    \caption{Sketch showing three different overlap degrees of particles \textbf{A} and \textbf{B}. The Overlap Ratio (OLR) is defined as $L/2R_\mathrm{B}$.}
    \label{fig:OLR}
\end{figure}

Further discussion of the processing algorithm and subsequent validation requires some definition of overlap degree and for this the Overlap Ratio (OLR) has been introduced. The overlap ratio is pictorially depicted in Fig.~\ref{fig:OLR} and is defined as $L/2R_\mathrm{B}$. An overlap ratio of OLR$\le 0$ indicates no overlap and an OLR $> 1$ designates complete overlap. 

\begin{figure}[h]
    \centering
    \includegraphics[width=0.6\linewidth]{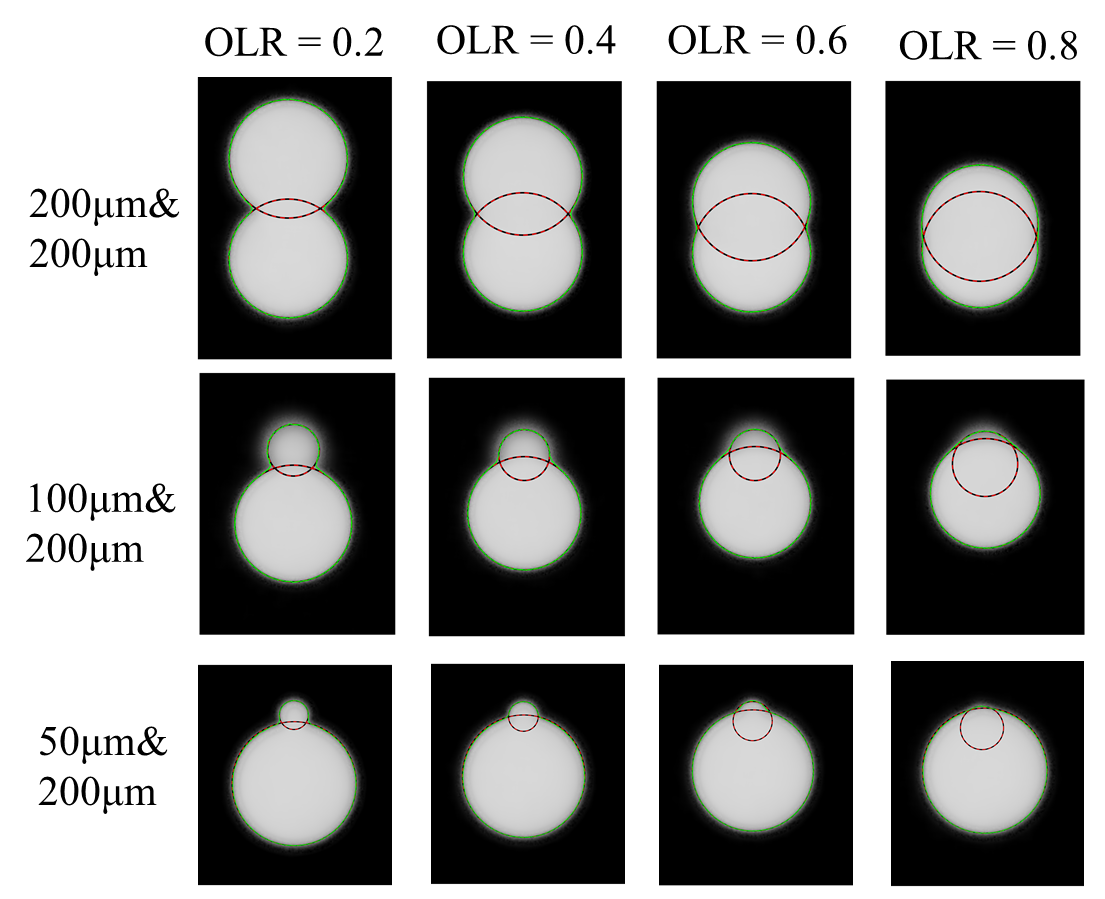}
    \caption{Example images with varying overlap ratios of two particles. Particle \textbf{A} has a diameter of 200~$\mu$m. Particle \textbf{B} varies in diameter between 50~$\mu$m and 200~$\mu$m. Example images for four overlap ratios are shown.  The green circles give the contour at $g_\mathrm{t}=0.5$ and the red circles are the circles fitted to the contour portion used for averaging the  gray-level gradient. In these examples the particles have different $z$ positions; hence different blur kernels. This will be discussed in subsection~\ref{subsec:validation multiple}.}
    \label{fig:overlap examples}
\end{figure}

To illustrate the procedure, a number of overlapping images are shown in Fig.~\ref{fig:overlap examples}.  In this figure the overlap ratio has been systematically varied for three example cases of particle size combinations. These images were acquired using the apparatus to be discussed in subsection~\ref{subsec:experimental apparatus}. 
\subsection{Procedure for Non-Spherical Particle Images}
\label{subsec:Non_Spherical Particles}
The above procedure is for spherical particles with a circular projected image. For non-spherical particles, this procedure is  slightly modified. The contour of the non-spherical particle is again determined at the selected $g_\mathrm{t}$ value using subpixel resolution. The diameter of an area equivalent circle ${d_\mathrm{t}}^{\prime}$ is then computed and this diameter is used, together with the average  gray-level gradient, to compute the dimensionless parameter $\tilde{\rho}_\mathrm{t}$, and to compute the size ${d_o}^{\prime}$ and location of a spherical particle which would have generated the area equivalent circular image. The ratio of the computed size ${d_o}^{\prime}$ to the diameter of the area equivalent circle ${d_\mathrm{t}}^{\prime}$ yields a scaling ratio, which is equal to the value $1/2\tilde{\rho}_\mathrm{t}$.
 The scaling ratio from this procedure is then applied to the $g_\mathrm{t}$-contour of the non-spherical particle to obtain an estimate of the true particle shape and size. The particle position along the $z$ axis is simply taken as the position of the approximating spherical particle. For this re-scaling step a procedure has been coded in which a pseudo-center of the particle is first computed, being  the  average of all contour coordinates in the $x$ and $y$ directions. The scaling ratio is then applied to the distance from the computed center to the respective contour coordinates and this new distance becomes the contour of the true particle shape (times the magnification factor) in the image plane.

\subsection{Related Work}
\label{subsec:related work}
It is clear from the algorithms description for the non-spherical particles and overlapping images that the blur kernel is not constant over the entire image. This is directly evident from Fig.~\ref{fig:gradient overlapping particles},   in which the change of blur kernel arises due to the difference in defocus degree. However, blur kernel changes also arise due to mutual gradient interaction in the case of overlapping images (contours b-d and f-h in Fig.~\ref{fig:gradient overlapping particles}) or from the non-circular shape  for the case of non-spherical particles. This situation is not unknown in image processing (\cite{junjie2023image}), where blur map evaluation based on gradient differences are then used in non-blind deconvolution. This avoids so-called boundary-ringing artifacts. For spherical particles with overlapping images this problem is circumvented by estimating the blur kernel parameter $\sigma$ from regions of low gradient differences (contours a-b and h-e in Fig.~\ref{fig:gradient overlapping particles}). For non-spherical particles, this effect is not avoided with the present algorithm.

\section{Validation  Experiments}
\label{sec:validation experiments}
\subsection{Experimental apparatus}
\label{subsec:experimental apparatus}
\begin{figure}[h]
    \centering
    \includegraphics[width=0.5\textwidth]{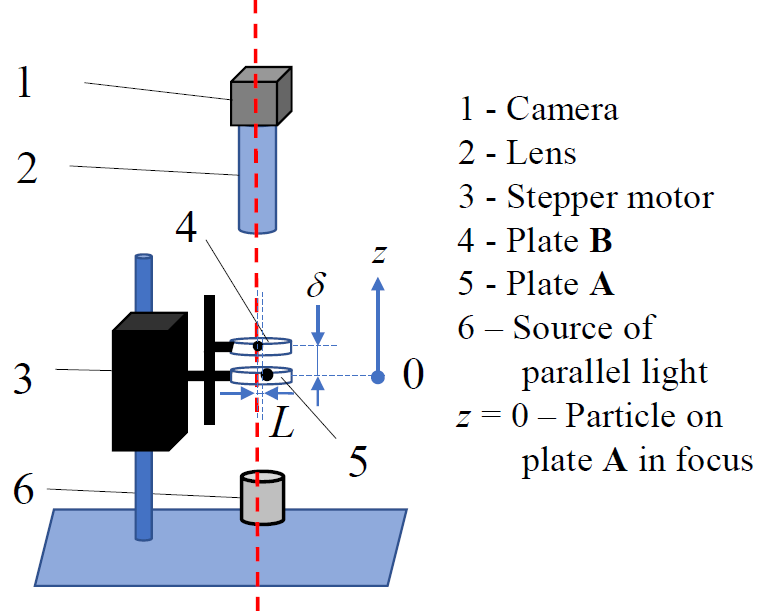} 
    \caption{Schematic diagram of the experimental apparatus for generating images with known particle sizes, positions and shapes.}
    \label{fig:experimental apparatus}
\end{figure}

To provide  images from particles with known size, position, shape and degree of overlap, the apparatus schematically pictured in Fig.~\ref{fig:experimental apparatus} was used. With this apparatus either one or two sample plates can be mounted. The distance between the plates ($\delta$) can be manually adjusted and the position of the plates can be traversed along the optical ($z$) axis using a stepper motor. Particles are placed on the  plates before mounting. In this manner both in-focus and out-of-focus images of the same particles can be acquired. The in-focus images allow the exact shape and size of all particles on the respective sample plate to be determined. The origin of the $z$-axis is chosen such that $z=0$ corresponds to the particle on plate \textbf{A} being in focus.  The value of the manually adjusted $\delta$ is determined from the stepper motor displacement between the particles on the sample plates \textbf{A} and \textbf{B} being in focus.   Thus, ground truth for both the particle size and position is always available to evaluate the accuracy of any DFD image processing algorithm.

To put the range of changes in $z$ or  $\delta$  expressed in millimeters into context, the depth of field for the optical system can be used for comparison. The depth of field is given approximately as
\begin{equation}
    \label{eq:dof}
    \mathrm{DOF} \approx 2 c f^\# \left( \frac{z_0}{f}-1 \right)
    \end{equation}
where $c$ is the circle of confusion, taken here as 0.03~mm, $f^\#$ is the f-stop (20.8), $f=53$~mm is the focal length of the lens, and the standoff distance $z_0=86$~mm.
This yields  DOF $\approx 0.74$~mm.

\subsection{Validation  for  particles with overlapping images}
\label{subsec:validation multiple}
Validation of the analysis when overlapping images are acquired is performed using the apparatus described above in subsection~\ref{subsec:experimental apparatus} and using the images similar to those  already shown in Fig.~\ref{fig:overlap examples}. These images were generated by using  calibration plates with varying dot sizes. Images were acquired for the three particle size combinations shown in Fig.~\ref{fig:overlap examples} with varying OLR and with two values of $\delta$, 0.1~mm and 0.4~mm. The intention of performing this systematic parameter variation is to provide a first qualitative impression of the maximum number concentration which can be reliably tolerated by the DFD technique using this analysis algorithm. 

\begin{figure*}[ht]
    \centering
     \includegraphics[width=1\linewidth]{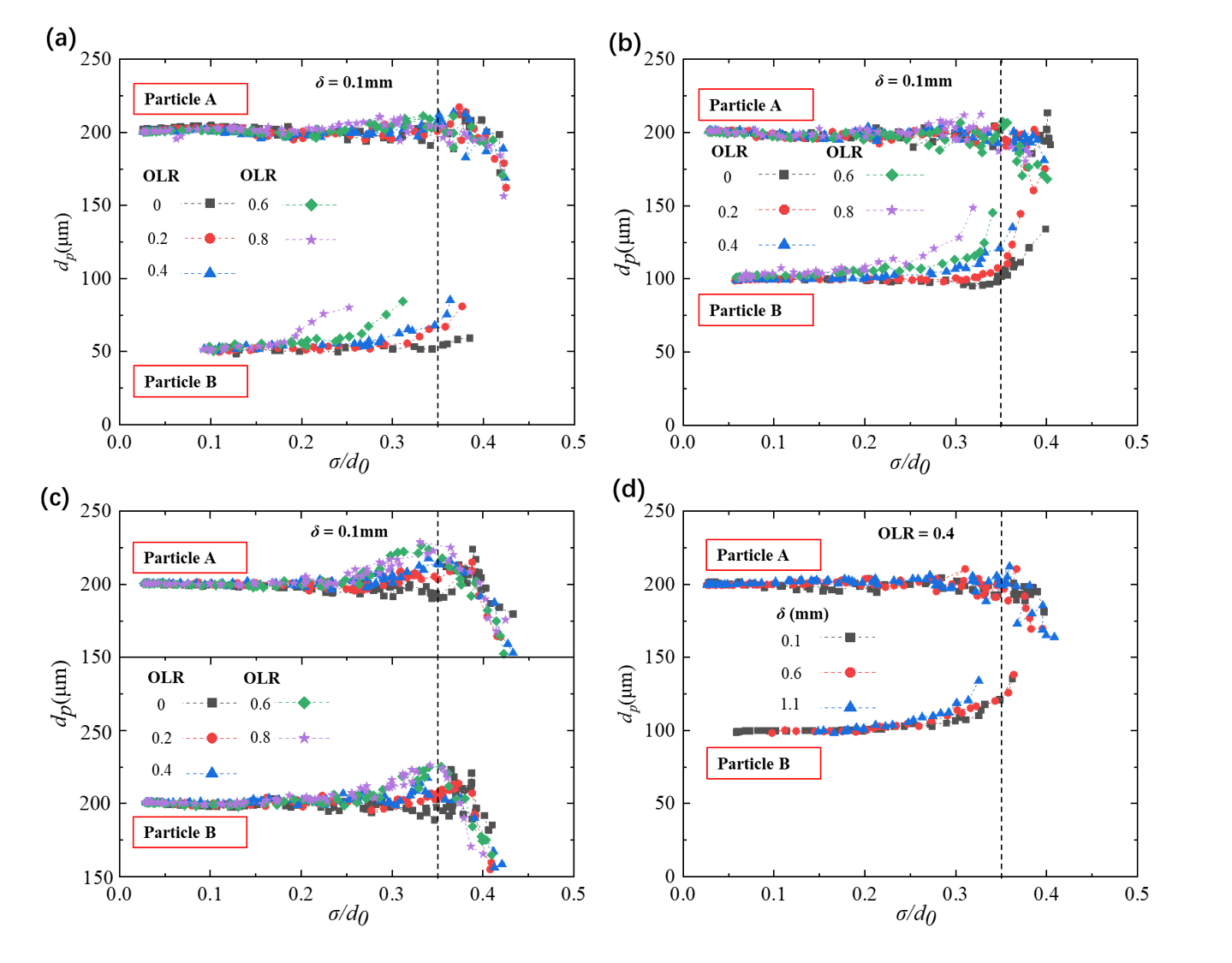}
    \caption{Parametric study in which the degree of blurring is varied by changing $z$. a), b) and c) uses a Particle \textbf{A} diameter of 200~$\mu$m, a $\delta =0.1$~mm and a Particle \textbf{B} diameter of 50~$\mu$m, 100~$\mu$m and 200~$\mu$m respectively. d) The OLR is held constant at 0.4 and $\delta$ is varied using the particle diameter combination 100~$\mu$m and 200~$\mu$m. The vertical black dashed lines marked on the graphs indicate   $\widetilde{\sigma}=0.35$, corresponding to the expected limitation of the single camera DFD technique. }
   \label{fig:parametric study}
\end{figure*}

The results of this parametric study are shown in Fig.~\ref{fig:parametric study}. In each of the four diagrams shown in this figure, the degree of blurring was systematically varied by changing $z$. 
For values of $z=0$  Particle \textbf{A} is in focus and if $\delta =0$, Particle \textbf{B} is also in focus. The validity limit of the processing algorithm, i.e., $\widetilde{\sigma}=0.35$, is shown in each diagram as a vertical dashed line.

It is clear that in all cases of  OLR = 0, the size evaluation is accurate up to the validity limit (and sometimes beyond). This is perhaps not intuitively obvious, since even with no overlap, the blur of each particle is mutually affecting the image of the other particle. However, the proposed algorithm eliminates these regions of mutual interaction when estimating $\tilde \sigma$. Furthermore, it appears that an overlap ratio of 0.4 appears to be tolerable, while still maintaining a size accuracy to within approx. 5\%. Above this value of OLR, size measurements can still be performed, albeit only to within a lower range of $\Tilde{\sigma}$ (degree of out-of-focus), especially for the smaller particle.

Restricting the OLR to 0.4 and below, the size estimate remains very accurate even with substantial values of $\delta$, as indicated in Fig.~\ref{fig:parametric study}d. Finally, it is instructive to compare the range over which reliable results can be achieved to the nominal DOF of the optical system (0.74~mm). For instance, for the 100~$\mu$m particle, a value of $\sigma/d_0=\tilde\sigma=0.35$ corresponds to approx. 2.95~mm, for the 200$\mu$m particle approx. 5.9~mm.  This underlines how effective the DFD technique is in quantifying particle size far out of the depth of field.

\subsection{Validation  for non-spherical particles}
\label{subsec:validation non-spherical}
\begin{figure}[ht]
    \centering
     \includegraphics[width=0.7\linewidth]{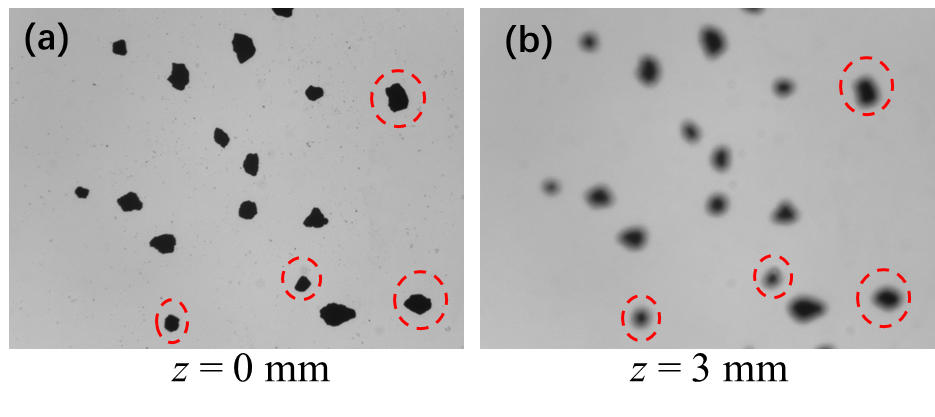}
    \caption{Images of coal particles placed on a glass plate: (a) in focus; (b) out of focus.}.
    \label{fig:coal particles}
\end{figure}

Coal particles have been used to validate the above image analysis procedure for non-spherical particles. The coal particles have been placed on plate \textbf{A}. The image in the in-focus plane ($z=0$) provides ground truth with which all following results can be rendered dimensionless. Such an in-focus image of several coal particles is pictured in Fig.~\ref{fig:coal particles}a. An example of an out-of-focus image of the same particles is shown in Fig.~\ref{fig:coal particles}b. In this figure, four particles have been circled, which will be used to demonstrate the accuracy of the analysis.

\begin{figure}[ht]
    \centering
     \includegraphics[width=0.7\linewidth]{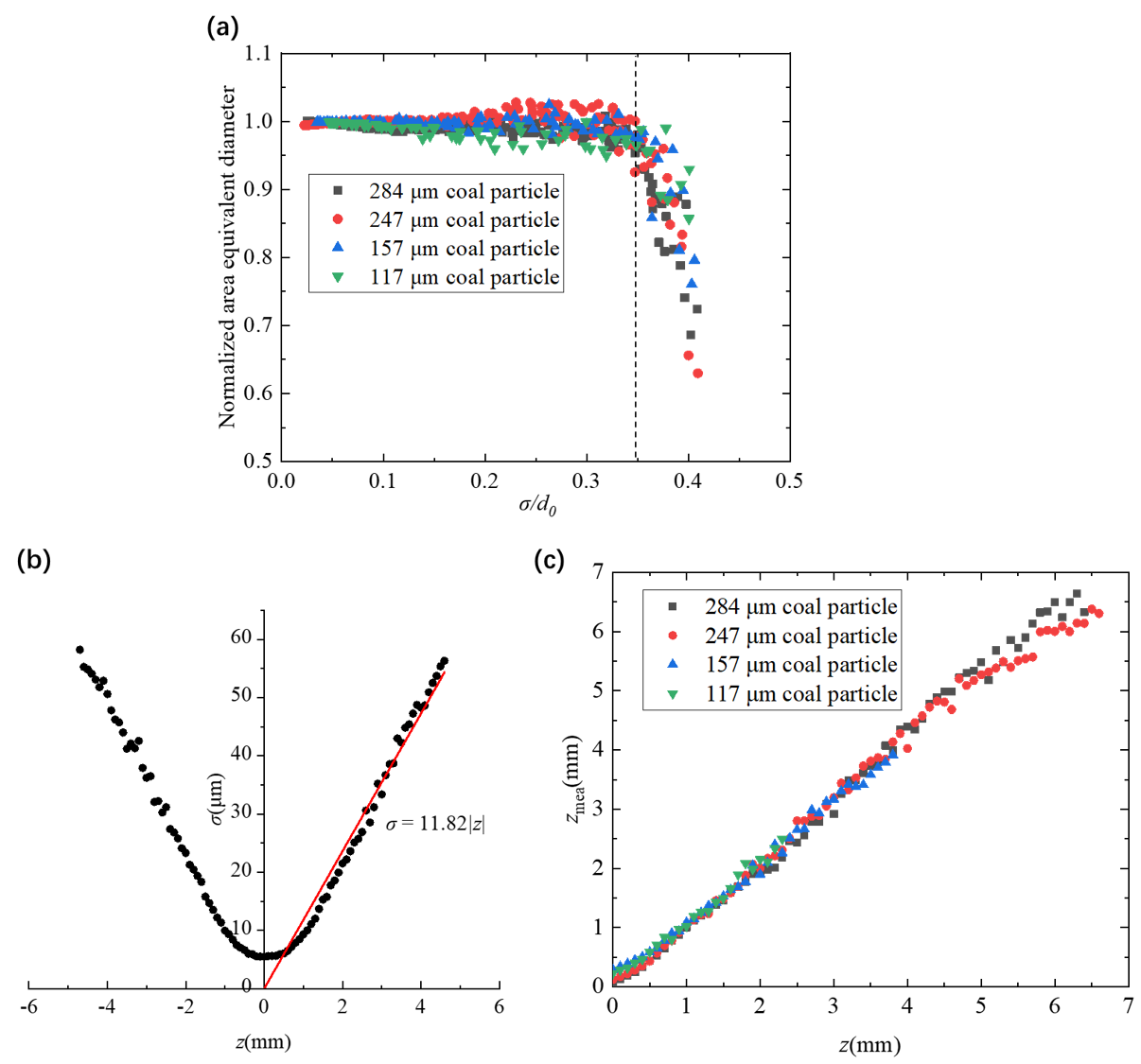}
    \caption{(a) Normalized area equivalent diameter of the four coal particles shown as a function of out-of-focus distance $z$. The vertical dashed line corresponds to $\widetilde{\sigma}=0.35$, considered the validity limit of the single camera DFD technique; (b) The computed value of $\sigma$ as a function of $z$. The portion of the curve exhibiting a linear slope relates to Eq.~(\ref{eq:sig-delz}); (c) The computed $z$ position of the four coal particles compared with the correct position. As a comparison, note that the depth of field of the optical system is 0.74~mm. }
    \label{fig:normalized coal particles}
\end{figure}

The result of analysing these four particles at a range of $z$ values over which the particle still resulted in gray values exceeding the limit set to $g_\mathrm{t}=0.5$ is shown in Fig.~\ref{fig:normalized coal particles}. Fig.~\ref{fig:normalized coal particles}a expresses the  area equivalent diameter 
normalized with the ground truth area equivalent diameter as a function of the $z$ displacement, expressed on the x-axis as the normalized blur standard deviation.  The validity range, given as $\widetilde{\sigma}=0.35$ is approximately constant for all particle sizes and has been shown as a vertical dashed line in Fig.~\ref{fig:normalized coal particles}a. The results indicate that the size of the particle is very well estimated, deviating less than 5\% within the validity range.

The relation between the $z$ displacement and the blur standard deviation is shown in Fig.~\ref{fig:normalized coal particles}b and, as expected is linear. The slope of this curve, either in front of ($z>0$) or behind ($z<0$) the in-focus plane corresponds to the apparatus constant $\beta$, as expressed in Eq.~(\ref{eq:sig-delz}). The blur kernel standard deviation does not go to zero when the particle is in focus. This residual blurring has two causes. For one, the pixel resolution limits the sharpness of the contour, even with sub-pixel interpretation, and second, there is a physical limit of diffraction. The pixel size of the detector is 3.45~$\mu$m; however, the subpixel resolution achieved in the image processing is considerably smaller. The Airy disk diameter, representing the diffraction limit of the optical system, is approx. 6.7~$\mu$m, which coincides very closely with the lower limit of resolution indicated in Fig.~\ref{fig:normalized coal particles}b. This resolution for small defocus distances can be improved by either increasing the object plane distance from the lens, resulting in a smaller magnification, or by opening up the aperture, which results in a smaller DOF. This indicates that there are compromises to be made in laying out the optical system. 

 Fig.~\ref{fig:normalized coal particles}c indicates that also the $z$ position of the particle can be reliably estimated. Here the computed particle position is compared with the known, true particle position. 

\begin{figure}[h]
    \centering
     \includegraphics[width=\linewidth]{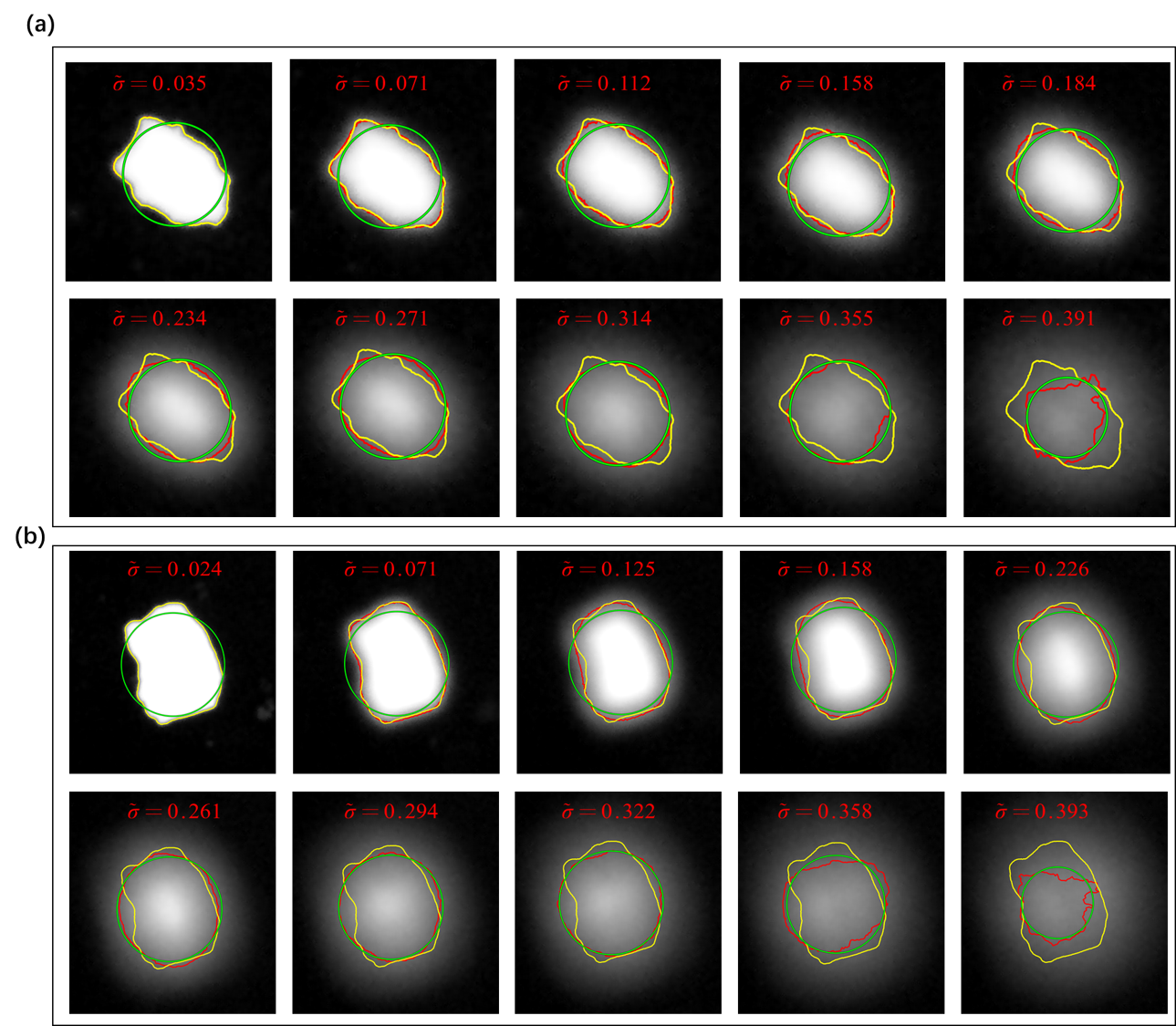}
    \caption{ Image processing results of coal powder particles at different values of $\widetilde{\sigma}$, i.e. related to $z$ (see Fig.~\ref{fig:normalized coal particles}b). The yellow line represents the contour of the particle in focus, the red line represents the reconstructed boundary, and the green line represents the equivalent area circle obtained according to the DFD algorithm: (a) 284~$\mu$m coal particle; (b) 247~$\mu$m coal particle, whereby these diameters refer to the area equivalent diameter.  }
   \label{fig:Non spherical Particle}
\end{figure}

The above validation was for size and position of the non-spherical particle and Fig.~\ref{fig:Non spherical Particle} explores the possibility of also reconstructing the shape of the particle. In this figure the particle is systematically shifted out of focus (increasing $\sigma$) and the reconstructed shape (red contour) is compared with the true shape (yellow contour) and with the area equivalent circle computed in estimating the particle size (green circle). These images indicate clearly that there is a  loss of shape information as the particle moves out of focus. Effectively the Gaussian blur kernel is acting as a low-pass filter generating a low resolution (LR) image from a high resolution (HR) image. As mentioned in subsection~\ref{subsec:related work}, there have been recent advances in reconstructing HR images using deblurring deconvolution, possibly in combination with deep learning. These algorithms  allow much improved feature extraction (\cite{amrollahi2023image,dong2021dwdn}). This is a subject going beyond the scope of the present study, but one which is being actively pursued by the authors. Nevertheless, in these examples reasonable shape estimates can be made without any further enhancements up to approx. $\tilde{\sigma}=0.25$.

\section{Discussion and Conclusions}
\label{sec:Conclusions}
This study has demonstrated two advancements to the Depth from Defocus technique for characterising dispersed, two-phase flows. In both cases  a novel approach of estimating the Gaussian blur kernel arising from the defocusing is invoked, utilizing the average gray-level gradient around the contour of the image.  The first advancement concerns the situation encountered with higher number densities of the dispersed phase. Under these conditions images of neighbouring particles will overlap, necessitating modified processing algorithms. In the present case again a gradient based deblurring has been employed, whereby the blur kernel is directly derived from the gray-level gradient, assuming a spherical particle. The blur gradient is monitored around the composite image and the blur kernel is derived based on regions of relative constant gradient values. This allows for different blur kernel values for each imaged particle. Although not yet investigated, this approach should be extendable to clusters with more than two  particles. 

The results show that accurate results can be achieved with overlap ratios up to OLR=0.4 at dimensionless blur kernel standard deviations of 0.35. This latter value can be considered a practical limit for gradient based blur kernel estimation. This limitation of overlap ratio does not translate easily into a direct limiting number density, but it does act as a good measure to monitor when performing measurements, signalling when tolerable number density limits have been reached. Future work will employ a Monte Carlo simulation to explore number density limits based on this information.

The second advancement presented in this study concerns the measurement on non-spherical particles, resulting in non-circular blurred images. The size and position of the non-spherical particle is determined using an area equivalent circular contour, yielding excellent agreement with the known ground truth values. The shape of the particle was also estimated using a simple re-scaling of the gray level contour, using a simple re-sizing procedure. The shape estimate was less accurate, showing clear signs of boundary-ringing where higher wavenumber components in the contour were present. Future work will explore recent advances in non-blind deblurring techniques involving deconvolution combined with neural networks (deep learning). Nevertheless, accurate estimates of number and volume density are still expected, since both size and location of the non-spherical particles are accurately estimated.

\section*{Acknowledgments}


\bibliographystyle{elsarticle-harv} 


%
%

\bibliography{elsarticle-harv}

\end{document}